\documentclass{aa}
\usepackage{graphicx}
\usepackage{txfonts}
\newbox\blankbox
\def\blank{\setbox\blankbox=\hbox{0}\kern\wd\blankbox}
\begin{document}
%
 \title{Variable stars in the globular cluster M\,13}
 \author{G.~Kopacki, Z.~Ko\l{}aczkowski, and A.~Pigulski}
 \institute{Wroc\l{}aw University Observatory,
            Kopernika 11, 51-622 Wroc\l{}aw, Poland}
 \date{Received .....; accepted .....}
 \offprints{G.~Kopacki,\\ e-mail: {\tt kopacki@astro.uni.wroc.pl}}
 \abstract{%
  Results of a search for variable stars in the
  central region of the globular cluster M\,13 are presented.
  Prior to this study, 36 variable and suspected variable stars 
  were known in this cluster 
  (Osborn \cite{osborn00}; Clement et al.\ \cite{clementetal01}).
  Of these stars, five were not observed by us.  
  We find v3, v4, v10, v12, and v13 to be constant in light.
  Surprisingly, only two out of the ten variable star candidates
  of Kadla et al.\ (\cite{kadlaetal80}) appear to be variable.
  Both are RRc variables.
  Additionally, three RR Lyrae stars and
  one SX Phoenicis variable are discovered. 
  Three close frequencies are detected for an RRc
  star v36. It appears that this variable
  is another multi-periodic RR Lyrae star pulsating
  in non-radial modes. 
  Light curves of the three known BL Herculis stars and
  all known RR Lyrae stars are presented.
  The total number of known RR Lyrae stars in M\,13 is
  now nine. Only one is an RRab star.
  The mean period of RRc variables amounts to
  $0.36\pm0.05$ d, suggesting that M\,13 should be included
  in the group of Oosterhoff type II globular
  clusters.
  Mean $V$ magnitudes and ranges of
  variation are derived for seven RR Lyrae 
  and three BL Herculis variables.
  Almost all observed bright giants show some
  degree of variability. In particular, we confirm
  the variability of two red giants announced to be variable 
  by Osborn (\cite{osborn00}) and in addition
  find five new cases. 
 }           

 \maketitle

 \section{Introduction}

 This is the third paper in a series devoted to
 the CCD photometry of variable stars in bright globular clusters
 of the northern sky.
 In the first two papers (Kopacki \cite{kopacki00},
 \cite{kopacki01}) the results of the
 search for variables in M\,53 and M\,92 were presented.
 Application of the image subtraction
 method 
 allowed us to confirm variability of 11 RR Lyrae stars and
 discover ten new variables:
 eight RR Lyrae and two SX Phoenicis stars.
 In the present paper we give the results of the photometric study
 of yet another cluster, M\,13. An announcement
 of this work was given by Kopacki et al.\ (\cite{kopackietal02}).
 
 M\,13 (NGC\,6205, C1639+365) is an intermediate
 metallicity globular cluster. Its metallicity on 
 the Zinn's (\cite{zinn85}) scale is [Fe/H] = $-1.65$.
 The cluster is located relatively far from the Galactic plane.
 In consequence, its reddening is small; 
 $E(B-V)$ = 0.02 mag according to
 Schlegel et al.\ (\cite{schlegeletal98}).
 Many colour--magnitude diagrams of this cluster
 have been published. As far as we are aware,
 the most recent CCD photometric study of M\,13
 is that of Rey et al.\ (\cite{reyetal01}).
 All these works show that the horizontal branch 
 of M\,13 is predominantly blue. 
 
 In the next section we give an
 account of the previous work on variable stars
 in M\,13. In subsequent sections we present 
 our observations and the results of our search for variable stars
 in the central region of the cluster using
 the Image Subtraction Method of
 Alard \&\ Lupton (\cite{alardlupton98}).
 We also comment on variability of all 
 observed suspected variable stars.
 For stars proved by us to be variable we determine periods
 and plot the light curves.

 \section{Variable stars in M\,13}

 The latest version of the Catalogue of Variable Stars in
 Globular Clusters (CVSGC, Clement et al.\ \cite{clementetal01})
 lists 33 variable and suspected variable stars in M\,13. Hereafter,
 the numbering system of this catalogue will be used.
 There are, however, other stars in M\,13 suspected to
 be variable but not included in the CVSGC. In these
 cases the numbering scheme of Ludendorff (\cite{ludendorff05}) 
 will be used, and star numbers of this system will be prefixed with
 the letter `L'.
 
 
 Below we give chronological description
 of variable star history in M\,13. One should note,
 however, that the numbering system of the CVSGC
 does not follow exactly the discovery order.

 Altogether, three Cepheids of the BL Herculis type 
 (v1, v2, and v6) and four RR Lyrae stars
 (v5, v7, v8, and v9) are known in the cluster. 
 The remaining variable stars are mainly semiregular 
 red giants. 

 The first variable stars in M\,13, v1 and v2, were discovered
 in 1898 by Bailey (\cite{bailey02}). 
 Barnard (\cite{barnard00a}, \cite{barnard00b}), 
 after hearing about the findings of Bailey, 
 rediscovered v2 and determined 
 its period to be 5.1 d. For v1 Barnard (\cite{barnard09}) 
 derived a period of 6.0 d. It appeared that both these
 variables were Cepheids.
 The next variable star to be discovered in the cluster, now
 known as v7, was found by Barnard (\cite{barnard14}). 
 He also reported L258 and L682 as probable variable stars.
 The variability of v7 was confirmed by Shapley (\cite{shapley15b}).
 Subsequently, Shapley (\cite{shapley15a}) identified four
 additional stars (v3$\,-\,$v6) as variable, for which he 
 suspected short periods.  
%
%
%

\begin{figure*}
  \hbox to\hsize{\hss\includegraphics{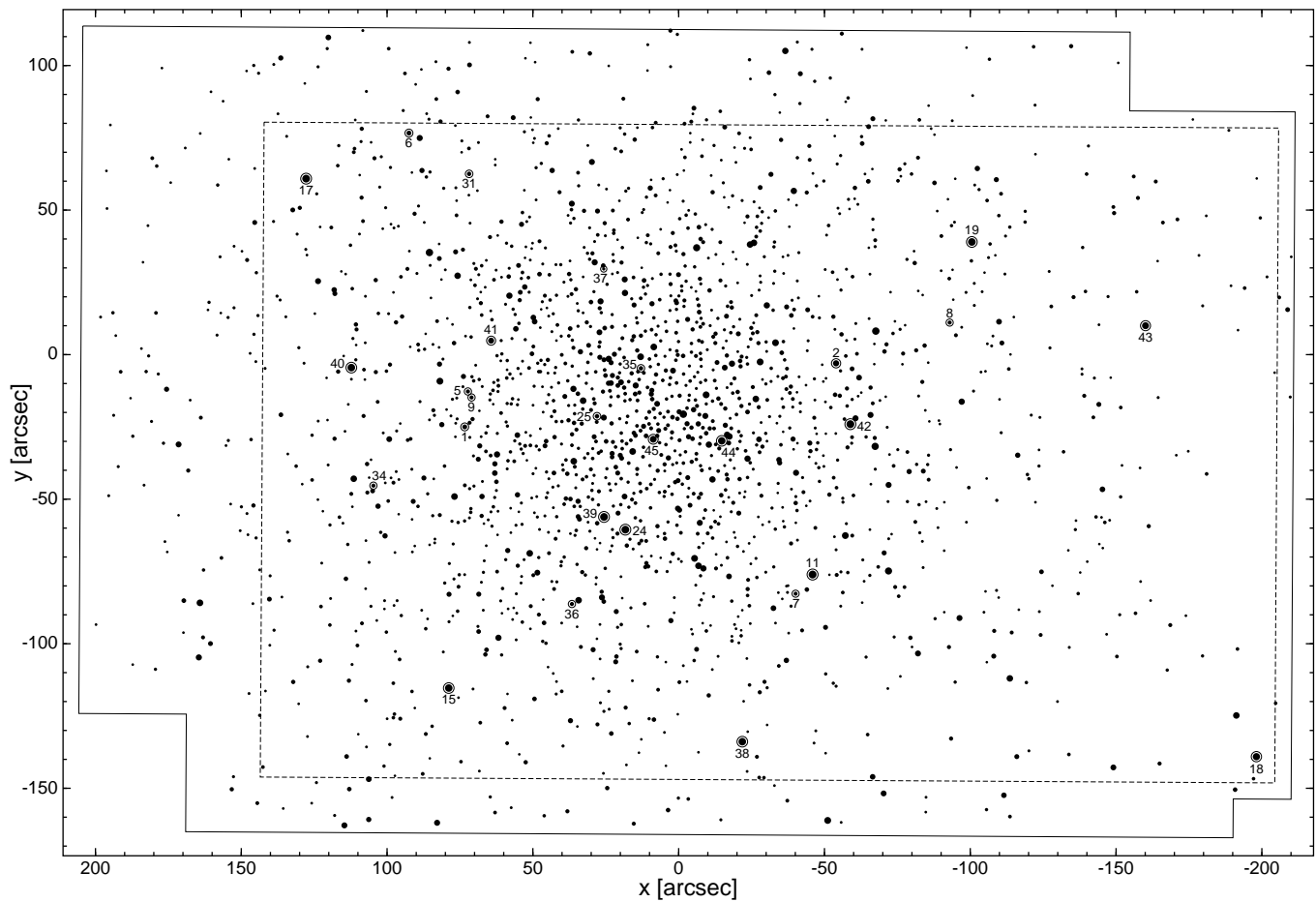}\hss}    
  \caption{Schematic view of the observed field of M\,13.
  The rectangle drawn with dashed lines denotes
  the region which was searched
  for variable stars using the image subtraction method.
  For clarity, only stars brighter than 17 mag in $V$ are shown.
  All observed variable stars are marked with open circles
  and are labeled with their numbers given in Tables~\ref{TabVarPos}
  or \ref{TabRedPos}.
  Positions $(x,y)$ are in the reference
  frame of the CVSGC. North is up, East to the left.}
  \label{FigFieldAll}%
 \end{figure*}

 The earliest version of the CVSGC (Sawyer \cite{sawyer39})
 listed seven variable or suspected variable stars in M\,13, v1$\,-\,$v7.
 Sawyer (\cite{sawyer40}) found four additional variables, v8$\,-\,$v11. 
 She could not resolve v9 and v5 (see Fig.~\ref{FigFieldAll}). 
 In the following paper, Sawyer (\cite{sawyer42})
 gave light curves for four variables and classified
 v1, v2, and v6 as Cepheids, and v8 as an
 RR Lyrae variable. She confirmed Barnard's (\cite{barnard09}) period of
 v2 and determined a correct period of 1.45899 d for v1.
 For the third Cepheid, v6, she derived a period of 2.11283 d.
 Although she could not resolve v5 and v9, 
 she concluded that these two variables are
 probably RR Lyrae stars. As to the suspected variables v10 and v11, 
 Sawyer (\cite{sawyer42}) considered them to be
 bright irregular variables.

 Kollnig-Schattschneider (\cite{kollnig42}) 
 extended the list of variable stars in M\,13 with
 three new suspects, v12$\,-\,$v14.
 For 11 variables (v1$\,-\,$v8, v12$\,-\,$v14)
 she derived periods and presented light curves. Although
 she classified v3, v4, v12, v13, and v14 as RR Lyrae stars
 with unusually short periods in the range $0.13-0.24$ d, 
 inspection of her light curves for these stars hardly indicates any
 variation. She determined also a period of about 0.75 d for an
 RRab variable v8.
%
%

 The next variable star in M\,13, v15, was
 found by Arp (\cite{arp55a}, \cite{arp55b}).
 In the first paper, Arp (\cite{arp55a})
 showed light curves of v1, v2, and v6,
 in the second (Arp \cite{arp55b}) he published light curves for 
 two certain RR Lyrae stars, v7 and v8. 
 For v7 Arp (\cite{arp55b}) derived a very short period of 0.2388 d.  
 The second edition of the CVSGC by Sawyer (\cite{sawyer55}) 
 lists 15 objects in M\,13, v1$\,-\,$v15. 
 However, variability of v3, v4, v12, v13, and v14 was questioned  
 on the basis of unpublished material of Arp and Sawyer.
 Tsoo (\cite{tsoo67}) found two suspected variable stars, 
 of which only one, v16, was included in the CVSGC.

 Osborn (\cite{osborn69a}) 
 studied period changes of v1, v2, v6, and v8.
 He also determined periods for
 two RR Lyrae stars, v5 and v9 (Osborn \cite{osborn69b},
 \cite{osborn73}).
 Demers (\cite{demers71}) presented $UBV$ photographic
 photometry of v1, v2, v6, v7, v8, and v11. He established
 that v11 is a long-period variable with period equal to
 92.5 d.
 Ibanez \&\ Osborn (\cite{ibanez73}) 
 corrected Arp's (\cite{arp55b}) period of v7 to 0.312929 d.
 Furthermore, Osborn \&\ Ibanez (\cite{osbornibanez73}) 
 reobserved v16 and found this star to be constant.
 Osborn et al.\ (\cite{osbornetal73}) examined new photographic observations of 
 v3 and concluded that this star is most likely non-variable.
 Russev (\cite{rusev73}) investigated eight variable stars in M\,13, 
 v1, v2, v5$\,-\,$v8, v11, and v15. He derived a period of 91.77 d for v11 and
 a period of 140.3 d for v15. Moreover, he detected variations
 in three red giants, L334, L414, and L917. 

 In the third edition of the CVSGC, Sawyer-Hogg (\cite{sawyerhogg73})
 listed 16 variables in M\,13, v1$\,-\,$v16. 
 In a paper on infrared photometry of M\,13,
 Russev (\cite{rusev74}) mentioned the variability 
 of one red giant, v17 (L973). His observations of 
 L414 indicated that this star is indeed variable.
 Subsequently, Fuenmayor \&\ Osborn (\cite{fuenmayor74}) 
 confirmed the variability of v17
 and derived a period of about 39 d, but did not
 detect any light variation of L414.
 In their photometric study of the central region of M\,13,
 Kadla et al.\ (\cite{kadlaetal76}) 
 found some evidence of variability for three stars,
 L194, L598, and L222; the first two 
 were subsequently designated as v19 and v24.
%
%

 The next variable red giant in M\,13, v20 (L70),
 was found by Pike \&\ Meston (\cite{pikemeston77}) 
 in their photometric study of the cluster 
 pulsating stars.
 At about the same time, Osborn \&\ Fuenmayor (\cite{osbornfuen77})
 presented photographic $B$ photometry for v2, v10, v11,
 and v15, as well as for two suspected red variables,
 v17 and L414. They showed that v11, v15, and v17 are
 semiregular variables with mean periods of 92.42,
 39.23, and 39.14 d, respectively, but could
 not confirm the variability of v10 and L414.
 Bisard \&\ Osborn (\cite{bisardosborn77}) 
 modified the period of the RR Lyrae star v7 to 0.3126626 d.

 Meinunger (\cite{meinunger78}) 
 announced finding three suspected variable stars.
 Russev \&\ Russeva (\cite{rusevruseva79a}) 
 analyzed archival and new observations of v17 and determined
 a period of about 43 d for this red giant.
 The same authors, Russev \&\ Russeva (\cite{rusevruseva79b}), 
 investigated once again the variation of v11 and concluded
 that its light curve is probably subject to periodic
 changes.
 A more thorough study of v1, v2, v4, v6$\,-\,$v8, v11, v12, and v15
 by Russev \&\ Russeva (\cite{rusevruseva79c}) 
 resulted in deriving periods of 0.298827 and 5.21753 d for
 v4 and v12, respectively, the two stars previously regarded
 constant (Sawyer-Hogg 1973). These authors suggested 
 that v4 is an RRc variable.
%

 Additional three suspected variable red giants in M\,13,
 L72, L240, and L261, 
 were found by Russeva \&\ Russev (\cite{rusevarusev80}).
 Of these stars, L72 is listed in the CVSGC as v18.
 Russeva \&\ Russev (\cite{rusevarusev80}) confirmed 
 the variability of v19 and derived
 periods of 35.62, 39.23, 41.25, 44.48, and 64.26 d 
 for v10, v15, v18, v19, and v20, respectively. 
 Kadla et al.\ (\cite{kadlaetal80}) 
 searched the core
 of M\,13, and discovered ten new suspected variable stars.
 These variables were included in the CVSGC with
 designations v21$\,-\,$v23 and v25$\,-\,$v31 (v24 has been discovered
 earlier by Kadla et al.\ \cite{kadlaetal76}). 
 Because of their photometric parameters,
 variables v27$\,-\,$v31
 were supposed by Kadla et al.\ (\cite{kadlaetal80}) to be of the 
 RR Lyrae type.

 Subsequently, White (\cite{white81}) 
 reported the discovery of light variations
 of three asymptotic branch giants in M\,13, 
 L687, L773, and L961. 
 Five stars in the cluster were investigated for variability
 by Russeva et al.\ (\cite{rusevaetal82}).
 These authors found L66 and v24 undoubtedly
 variable. The new variable L66 was denoted in 
 the CVSGC as v32. For the remaining three stars, 
 L526, L761, and L1067, Russeva et al.\ (\cite{rusevaetal82}) 
 did not detect any variability. 
 The period changes of short-period variable stars
 in M\,13 were studied by Russeva \&\ Russev (\cite{rusevarusev83}).
 They suggested that M\,13 belongs to the Oosterhoff type I
 globular clusters.

 A large sample of bright red giants in M\,13,
 including all previously suspected to be variable, 
 was checked for variability by Welty (\cite{welty85}).
 The result was the confirmation of variation for v11, v17$\,-\,$v20,
 and the detection of light changes in a new suspected
 variable, v33 (L954), which was considered constant by
 White (\cite{white81}). 
 Recently, Osborn (\cite{osborn00})
 presented $UBVRI$ photometry for variable stars in M\,13.
 He found v3, v4, v10, v12, v13, v15, and v18 to be 
 most probably constant in light and detected variability in 
 two red giants, L629 and L940. Moreover, Osborn (\cite{osborn00}) checked 
 White's (\cite{white81}) three possible red variables (L687, L773, L961)
 and Russeva \&\ Russev's (\cite{rusevarusev80}) two other suspects (L240, L261)
 and found them constant in light.

 \section{Observations and reductions}

 The CCD observations presented here were carried out at the
 Bia\l{}k\'ow station of the Wroc\l{}aw University Observatory
 with the same equipment as that described by Kopacki \&\ Pigulski
 (\cite{koppig95}).
 A 6$\times$4 arcmin$^2$ field of view covering
 the core of the cluster
 was observed through $V$ and $I_{\rm C}$ filters
 of the Johnson-Kron-Cousins $UBV(RI)_{\rm C}$ system.
 The observed field was chosen to include as many
 as possible of the known variable stars in M\,13.

 The observations were carried out 
 on 23 nights between 2001 February 27 and August 1.
 In all, we collected 324 and 342 frames
 in the $I_{\rm C}$ and $V$ bands, respectively.
 Usually, the exposure times amounted to 300 s for
 the $V$-filter frames and 200 s for 
 the $I_{\rm C}$-filter frames.
 On most nights the weather was
 very good. On some nights, however, the sky brightness
 was rather high due to the bright Moon;
 on some others the sky transparency was
 affected by thin cirrus clouds.
 The seeing varied over a rather wide range,
 between 1.7 and 3.9 arcsec, with a typical value of
 2.4 arcsec.

 The pre-processing of the frames was performed in the usual way and
 consisted of subtracting bias and dark frames and
 applying the flat-field correction.
 Instrumental magnitudes for all stars in the field
 were computed using the DAOPHOT profile-fitting software
 (Stetson \cite{stetson87}).
 All images were reduced in the same way as described by
 Jerzykiewicz et al.\ (\cite{jerzykiewicz96}).
 We identified 3124 stars in the observed field.
 A finding chart for the monitored field
 is shown in Fig.~\ref{FigFieldAll}. 

 Our average instrumental $V$-magnitudes were
 transformed to the standard ones using the
 recent $BV$ photometric data of Rey et al.\ (\cite{reyetal01}).
 To do this, we obtained
 on two nights, 2002 February 13/14 and 14/15, 
 12 $I_{\rm C}$-filter and 11 $V$-filter CCD frames
 of an additional field shifted about 2.5 arcmin
 to the east in respect to the main observed field.
 Thirty six stars in common with Rey et al.\ (\cite{reyetal01})
 were found, and the
 following transformation equation was obtained:
 $$
  \begin{array}{rll}
   V-v& = +\hbox{0.074}\,(v-i)+\hbox{13.474},&
    \kern0.7cm \sigma=\hbox{0.027},\\
  \end{array}
 $$
 where uppercase letters denote standard magnitudes
 and lowercase letters, the instrumental ones;
 $\sigma$ is the standard deviation from the fit.

 Using the above given equation, instrumental $V$-filter 
 magnitudes of the variable stars
 were transformed to the standard system in the same way
 as described by Kopacki (\cite{kopacki01}).

 In order to search for variable stars in the core of the
 cluster we reduced our CCD frames using the Image
 Subtraction Method (ISM) package developed by Alard \&\ Lupton
 (\cite{alardlupton98}). This method allows obtaining
 very good quality light-curves even
 in very crowded fields and is 
 now in common use.
 The same procedure of reductions
 as in Kopacki (\cite{kopacki00}) was followed; we refer
 the reader to this paper for details.

 In the next step 
 for each variable-star candidate detected
 in the ISM variability map we computed the
 AoV periodogram of Schwarzenberg-Czerny (\cite{schwarz96})
 in the frequency range from 0 to 30 d$^{-1}$.
 To identify variable and constant stars we checked visually 
 these periodograms as well as light curves.

 \section{Results of the variability survey}

 \begin{table}
  \caption[]{Types of variability, 
   corrected positions ($x,y$) relative to the cluster centre,
   periods ($P$) and epochs of light-maximum 
   ($T_{\rm max}=\rm HJD_{\rm max}-2400000$) of 
   the short-period pulsating stars in the
   observed field of M\,13. Coordinates are given in the reference
   frame of the CVSGC.}
  \label{TabVarPos}
  \hbox to\hsize{\hfill\vbox{\tabskip=4pt 
  \halign{#\hfil\tabskip=15pt&#\hfil&\hfil#&\hfil#%
    &#\hfil&#\hfil\tabskip=4pt\cr
   \noalign{\hrule\vskip1.7pt}
   \noalign{\hrule\vskip3pt}
   Var& Type& $x$ [$\,{}^{\prime\prime}$]& $y$ [$\,{}^{\prime\prime}$]& $P$ [d]& $T_{\rm max}$ \cr
   \noalign{\vskip3pt\hrule\vskip3pt}
    v1&  BL Her&      73.43&  $-$24.96&  1.459033& 52000.120\cr
    v2&  BL Her&   $-$53.88&   $-$3.00&  5.110818& 51999.600\cr
    v5&  RR Lyr&      72.37&  $-$12.59&  0.38180&  52000.330\cr
    v6&  BL Her&      92.52&     76.65&  2.112918& 51999.812\cr
    v7&  RR Lyr&   $-$40.02&  $-$82.56&  0.31269&  51999.984\cr
    v8&  RR Lyr&   $-$92.78&     11.04&  0.750316& 51999.785\cr
    v9&  RR Lyr&      70.86&  $-$15.03&  0.39278&  52000.194\cr
   \noalign{\vskip3pt\hrule\vskip3pt}
    v25& RR Lyr&      28.05&  $-$21.19&  0.42956&  52000.195\cr
    v31& RR Lyr&      71.73&     62.48&  0.31930&  51999.995\cr
   \noalign{\vskip3pt\hrule\vskip3pt}
    v34&  RR Lyr&     104.69&  $-$45.25& 0.38933&  51969.139\cr
    v35&  RR Lyr&      13.09&   $-$4.88& 0.32003&  52000.003\cr
    v36&  RR Lyr&      36.93&  $-$86.03& 0.31584\hbox to 0pt{ $P_1$\hss}& 52023.214\cr
    &&&& 0.30441\hbox to 0pt{ $P_2$\hss}& 52023.158\cr 
    &&&& 0.33497\hbox to 0pt{ $P_3$\hss}& 52023.086\cr 
    v37&  SX Phe&      25.70&     29.62& 0.04941&  52000.002\cr
   \noalign{\vskip3pt\hrule}}
   }\hfill}
 \end{table}

 Out of the 33 variable and suspected variable stars listed in the
 CVSGC, five were outside the field investigated by us.
 These were v14, v16, v20, v32, and v33. Moreover, the field
 we observed included two suspected variable red
 giants of Osborn (\cite{osborn00}). We found
 v3, v4, v10, v12, and v13 to be constant in light.
 Our conclusion is in agreement with Osborn's (\cite{osborn00})
 less accurate photometry for these stars. 
 It should be mentioned that v4 was classified
 in the CVSGC as an RRc star for which 
 Russev \&\ Russeva (\cite{rusevruseva79c})
 derived a period of 0.298827 d. 

\begin{figure}
  \hbox to\hsize{\hss\includegraphics{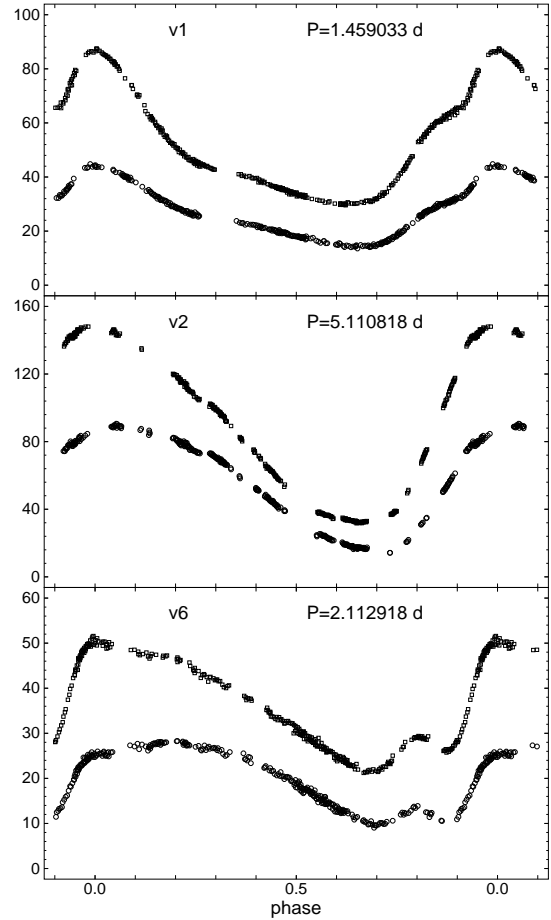}\hss}
  \caption{$V$-filter (squares) and
  $I_{\rm C}$-filter (circles) light-curves
  of the three BL Herculis variables.
  Ordinate is expressed in arbitrary flux units.
  Note the differences in the ordinate scales.}
  \label{FigLCWVIR}%
 \end{figure}

 For the three BL Herculis stars known in the cluster,
 v1, v2, and v6, light-curves with good phase coverage
 were obtained. 
 The $V$- and $I_{\rm C}$-filter light-curves of these three
 stars are plotted in Fig.~\ref{FigLCWVIR}.
 Our light-curves of v1 and v6 show clearly
 bumps on the ascending branches. This feature
 for v1 was previously suggested by  
 Osborn (\cite{osborn76} and references therein) and
 now is definitely confirmed.
 
\begin{figure}
  \hbox to\hsize{\hss\includegraphics{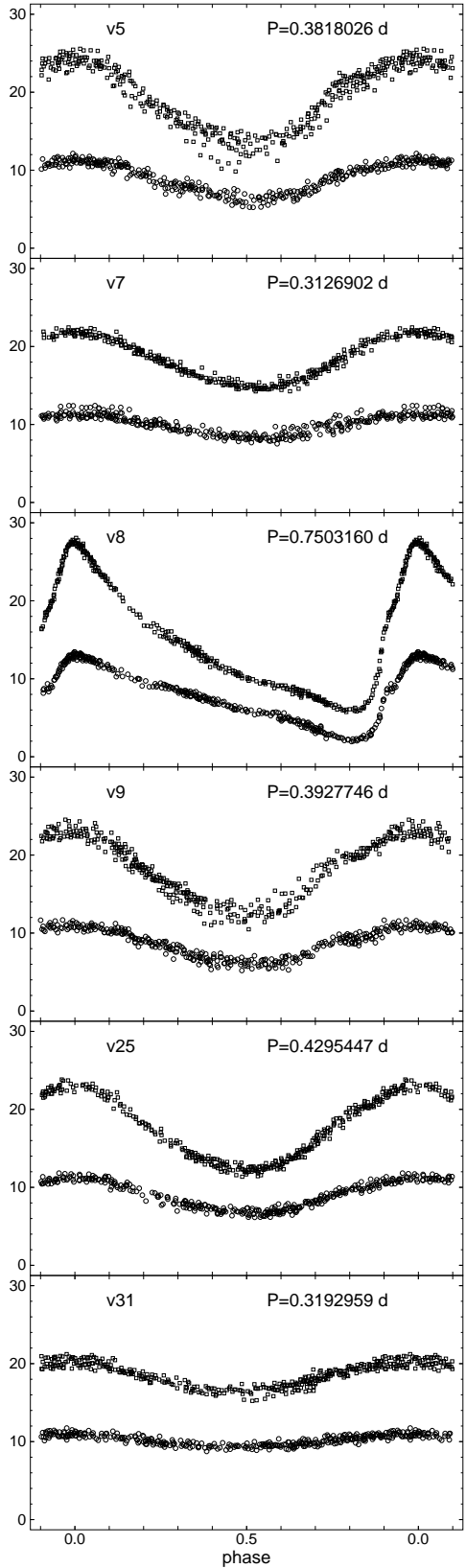}\hss}
  \caption{$V$-filter (squares) and
  $I_{\rm C}$-filter (circles) light-curves
  of the observed RR Lyrae stars
  listed in the CVSGC. The variables are arranged according
  to their numbers in that catalogue.
  Ordinate is expressed in the same flux units as in
  Fig.\ \ref{FigLCWVIR}.}
  \label{FigLCCVSGC}%
 \end{figure}

 Surprisingly, only two out of the ten suspected
 variable stars of Kadla et al.\ (\cite{kadlaetal80}),
 viz.\ v25 and v31, appear to be variable. 
 All eight remaining stars show 
 no evidence of variability in our data.
 We classify v25 and v31 as RRc variables.
 The $V$- and $I_{\rm C}$-filter light-curves 
 of these two RR Lyrae stars, as well as those of the four previously
 known variables of this type are shown in Fig.~\ref{FigLCCVSGC}.

 \begin{figure}
  \hbox to\hsize{\hss\includegraphics{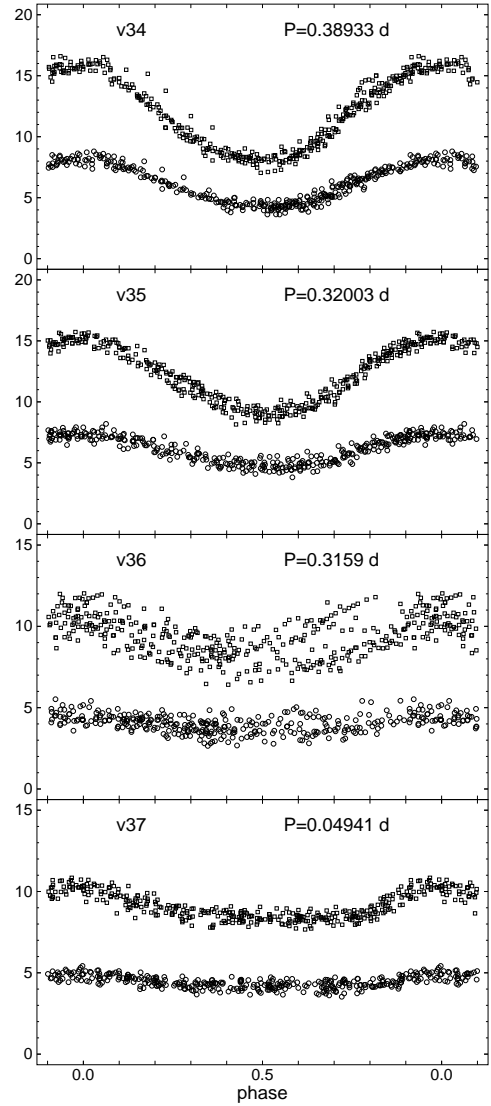}\hss}
  \caption{$V$-filter (squares)
  and $I_{\rm C}$-filter (circles)
  light-curves of four new variable stars:
  three RR Lyrae stars (v34$\,-\,$v36)
  and an SX Phoenicis star, v37.
  Ordinate is expressed in the same flux units as in
  Fig.\ \ref{FigLCWVIR}.
  Note the differences in the ordinate scales.}
  \label{FigLCKopRRAll}%
 \end{figure}

 In addition, four new short-period pulsating stars
 have been found in the cluster.
 Three of them are of RR Lyrae type,
 the fourth is an SX Phoenicis star.
 Extending the numbering scheme
 of the CVSGC for M\,13, we designate these new
 variables as v34 through v37.
 Their $V$- and $I_{\rm C}$-filter 
 light-curves are shown in Fig.~\ref{FigLCKopRRAll}.
 As can be seen in this figure,
 the scatter of observations in the light curve of
 the RRc star v36 is much larger than for
 other RRc variables. 
 This variable deserves more detailed
 analysis and will be discussed separately.

 Positions relative to the cluster centre,
 types of variability,
 adopted periods, and computed epochs
 of light-maximum of the short period variable 
 stars we observed are given in Table~\ref{TabVarPos}.
 Coordinates are given in the
 reference frame of the CVSGC, but were redetermined using positions of
 16 stars in the CVSGC. All these variables are also indicated in 
 Fig.~\ref{FigFieldAll}.

 The periods of the BL Herculis stars and the RRab variable
 v8 given in Table~\ref{TabVarPos} were taken from the 
 CVSGC. For all previously known RRc variables, v5, v7, and v9,
 we determined revised periods using our observations only. 
 These new periods
 produced smoother light-curves than the old ones.
 Other periods listed in Table~\ref{TabVarPos} were
 derived for the first time from our data alone
 using multi-harmonic AoV periodogram method of
 Schwarzenberg-Czerny (\cite{schwarz96}).

 \begin{figure}
  \hbox to\hsize{\hss\includegraphics{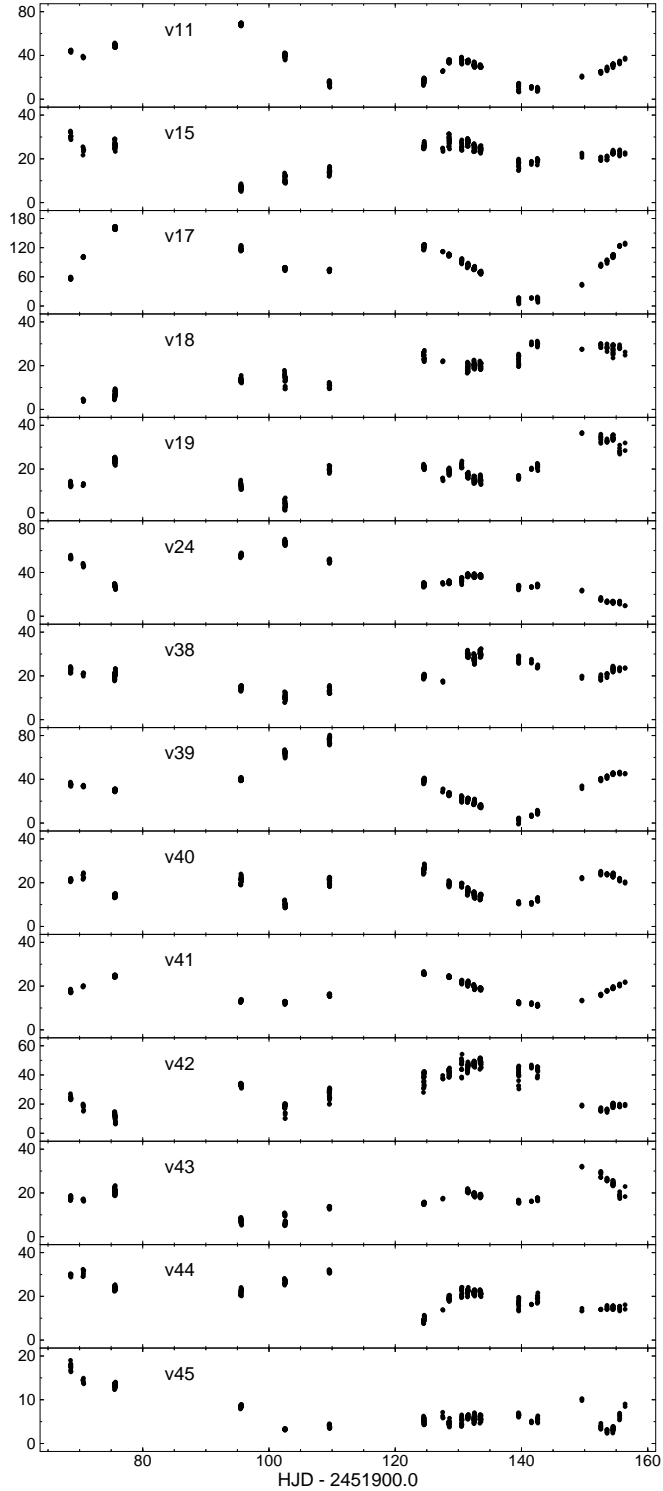}\hss}
  \caption{$V$-filter light-curves of 
  the observed variable red giants.
  Except v41, they all exhibit irregular
  brightness variations.
  Ordinate is expressed in the same flux units as in
  Fig.\ \ref{FigLCWVIR}.
  Note the differences in the ordinate scales.}
  \label{FigRedVar}%
 \end{figure}

 The total number of known RR Lyrae stars in M\,13 equals
 now nine. Only one of them is an RRab star.
 The mean period of RRc variables amounts to
 $0.36\pm0.05$ d, suggesting that M\,13 should be included
 in the Oosterhoff type II group of globular
 clusters.

 Bright giants in globular clusters usually
 show some degree of variability.
 In M\,53, for example, seven giants are known to be variable
 (Kopacki \cite{kopacki00}). On the other hand,
 no variable red giant was found by
 Kopacki (\cite{kopacki01}) in M\,92.

 \begin{figure}
  \hbox to\hsize{\hss\includegraphics{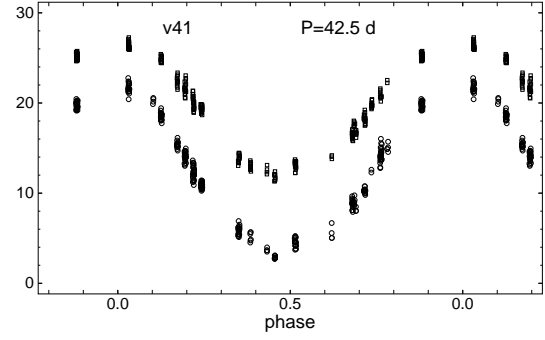}\hss}
  \caption{$V$-filter (squares) and
  $I_{\rm C}$-filter (circles) observations
  of the new variable red giant, v41, phased with
  the period of 42.5 d.
  Ordinate is expressed in the same flux units as in
  Fig.\ \ref{FigLCWVIR}.}
  \label{FigRedPer}%
 \end{figure}

 \begin{table}
  \caption[]{Positions ($x,y$) relative to the
   cluster center, average $V$-brightnesses,
   $\langle{V}\rangle$, and the ranges of variability,
   $\Delta V$, of the observed red-giant variables.
   Coordinates are given in the reference frame of the
   CVSGC. }
  \label{TabRedPos}
  \hbox to\hsize{\hfill\vbox{\tabskip=4pt 
  \halign{#\hfil\tabskip=15pt&\hfil#&\hfil#&\hfil#\hfil&
   \hfil#\hfil\tabskip=3pt\cr
   \noalign{\hrule\vskip1.7pt}
   \noalign{\hrule\vskip3pt}
   Var& $x$\hfil& $y$\hfil& $\langle{V}\rangle$& $\Delta{V}$\cr
   \noalign{\vskip1pt}
    & [$\,{}^{\prime\prime}$]\hfil& [$\,{}^{\prime\prime}$]\hfil&
    [mag]& [mag]\cr
   \noalign{\vskip3pt\hrule\vskip3pt}
    v11&      $-$45.81&  $-$76.07& 11.928& 0.13\cr
    v15&         78.88& $-$115.40& 12.139& 0.09\cr
    v17&        127.86&     60.97& 11.976& 0.38\cr
    v18&     $-$197.67& $-$138.87& 12.319& 0.11\cr
    v19&     $-$100.31&     38.88& 12.066& 0.09\cr
    v24&         18.46&  $-$60.42& 12.005& 0.24\cr
   \noalign{\vskip3pt\hrule\vskip3pt}
    v38&      $-$21.69& $-$133.88& 12.118& 0.07\cr
    v39&         25.63&  $-$56.05& 11.980& 0.22\cr
    v40&        112.03&   $-$4.25& 12.075& 0.08\cr
   \noalign{\vskip3pt\hrule\vskip3pt}
    v41&         64.49&      4.92& 13.155& 0.11\cr
    v42&      $-$58.83&  $-$24.04& 11.940& 0.10\cr
    v43&     $-$159.92&      9.96& 12.470& 0.07\cr
    v44&      $-$14.74&  $-$29.64& --& --\cr
    v45&          8.89&  $-$29.27& --& --\cr
   \noalign{\vskip3pt\hrule}}
   }\hfill}
 \end{table}

 CVSGC lists ten suspected red variables in M\,13, of which
 we observed seven. Except for v10, the other
 six above-mentioned red giants show light variations.
 We confirm also the variability of 
 the Russev's (\cite{rusev73}) suspect, L414,
 as well as the two
 red giants of Osborn (\cite{osborn00}), 
 L629 and L940. Since these three variables 
 are not included in the CVSGC, we name
 them with consecutive numbers as v38, v39, and v40,
 respectively.
 Moreover we discovered five new variable red giants
 which we designate v41 through v45.
 Positions of all red giants we found variable,
 computed in the same way as for RR Lyrae stars,
 together with the average $V$-filter brightnesses and
 ranges of variability
 are given in Table~\ref{TabRedPos}.
 Fig.~\ref{FigRedVar} shows the light curves of 
 these variables.
 All variable red giants are also indicated in 
 Fig.~\ref{FigFieldAll}.

 Almost all red variables seem to be semi- or
 irregular. In particular, we could not confirm 
 any period determined earlier for these variables
 (see Sect.~2). The only red giant 
 showing a periodic variation, at least
 over the time covered by our observations, is 
 v41. For this variable we derived a period
 of 42.5 d. Observations of v41,
 phased with this period, are presented
 in Fig.~\ref{FigRedPer}.
 
 The suspects not included in the CVSGC deserve a comment.
 Firstly, we identified three suspected variable stars
 of Meinunger (\cite{meinunger78}). Her star
 a is L765 and is outside the field we observed.
 Star b could not be found in the Ludendorff's
 list. In our CCD frames it forms a blend of two 
 faint stars situated close to L956. The equatorial
 coordinates of the brighter component are given in
 Table~\ref{TabPosEquat}. Both these stars are
 constant in light. The third suspect, star c, can be safely
 identified as L993, which is known UV-bright
 object in M\,13 (Zinn et al.\ \cite{zinnetal72}).
 On the basis of radial velocity and proper motion 
 measurements 
 (Zinn \cite{zinn74}, Cudworth \&\ Monet \cite{cudworthetal79}) 
 the star was regarded as a cluster member. 
 It shows no evidence of variability in
 our data.
 
 Of all other suspected variable stars in M\,13
 we observed the blue straggler L222 and two red
 giants, L261 and L334. We found these three stars
 constant in light.
 L222 also known as Barnard 29
 (Barnard \cite{barnard14}) 
 is a very interesting object, which 
 is now believed to be hot cluster 
 post-AGB star (Conlon et al.\ \cite{conlonetal94}).


 Altogether, the light curves have been obtained
 for 13 variable stars in the Cepheid instability strip
 and for 14 variable red giants. The observations are available
 in electronic form from CDS in Strasbourg via anonymous
 ftp to {\tt cdsarc.u-strasbg.fr} (or {\tt 130.79.128.5}).
 For each variable we give a list of 
 HJD of frame mid-exposure along with $V$ and
 $I_{\rm C}$ magnitudes and ISM fluxes, their errors,
 and air masses.

 Using 200 reference stars from the HST Guide Star
 Catalog, Version 2.2.01 we transformed our
 DAOPHOT rectangular coordinates
 of all observed variable and suspected variable stars in M\,13 into
 J2000.0 equatorial coordinates. These positions
 are given in Table~\ref{TabPosEquat}. 
 We found our coordinates shifted systematically
 $0.58\pm0.02$ arcsec east and $0.24\pm0.02$ south
 in respect to those derived by Osborn (\cite{osborn00})
 based on 39 stars in common. 
 Undoubtedly these differences result from Osborn's
 use of a different reference star catalogue, the USNO-A2.0.
 All objects
 under consideration were also cross-identified
 with stars in Ludendorff's (\cite{ludendorff05}) list.
 The Ludendorff's numbers are given in the second
 column of Table~\ref{TabPosEquat}.
 It should be noted that Kadla et al.\ (\cite{kadlaetal80}) 
 erroneously identified v28 as L569. Using their 
 coordinates for this star we found it to be in fact L519.

 \begin{table}
  \caption[]{Equatorial coordinates of all observed
   variable and suspected variable stars in M\,13. 
   L is a Ludendorff's number and `RG' denotes variable red giant.
   Stars a and b are Meinunger's suspects.}
  \label{TabPosEquat}
  \hbox to\hsize{\hfill\vbox{\tabskip=5pt 
  \halign{#\hfil\tabskip=18pt&\hfil#%
    &#\hfil&\hfil#\hfil&\tabskip=5pt&\hfil#\hfil\cr
   \noalign{\hrule\vskip1.7pt}
   \noalign{\hrule\vskip3pt}
   Var& L& Type& $\alpha_{2000}$& $\delta_{2000}$\cr
   \noalign{\vskip1pt}
     & & & [$^{\rm h}$ $^{\rm m}$ $^{\rm s}$]& [$^\circ$ $^\prime$ $^{\prime\prime}$]\cr
   \noalign{\vskip3pt\hrule\vskip3pt}
  v1&   816&  BL Her& 16 41 46.45& 36 27 27.7\cr
  v2&   306&  BL Her& 16 41 35.88& 36 27 48.3\cr
  v5&  806\hbox to0pt{$\beta$\hss}&  RR Lyr& 16 41 46.35& 36 27 39.9\cr
  v6&   872&  BL Her& 16 41 47.95& 36 29 09.6\cr
  v7&   344&  RR Lyr& 16 41 37.10& 36 26 28.8\cr
  v8&   206&  RR Lyr& 16 41 32.64& 36 28 02.0\cr
  v9&  806\hbox to0pt{$\alpha$\hss}&  RR Lyr& 16 41 46.24& 36 27 37.8\cr
  v11&   324&      RG& 16 41 36.61& 36 26 35.3\cr
  v15&   835&      RG& 16 41 46.98& 36 25 57.3\cr
  v17&   973&      RG& 16 41 50.89& 36 28 54.2\cr
  v18&    72&      RG& 16 41 24.06& 36 25 30.6\cr
  v19&   194&      RG& 16 41 31.98& 36 28 29.8\cr
  v24&   598&      RG& 16 41 41.91& 36 26 51.5\cr
  v25&   630&  RR Lyr& 16 41 42.68& 36 27 31.0\cr
  v31&   807&  RR Lyr& 16 41 46.25& 36 28 55.2\cr
  v34&   918&  RR Lyr& 16 41 49.06& 36 27 07.7\cr
  v35&   571&  RR Lyr& 16 41 41.43& 36 27 47.1\cr
  v36&    --&  RR Lyr& 16 41 43.45& 36 26 26.0\cr
  v37&    --&  SX Phe& 16 41 42.45& 36 28 21.8\cr
  v38&   414&      RG& 16 41 38.65& 36 25 37.7\cr
  v39&   629&      RG& 16 41 42.51& 36 26 56.0\cr
  v40&   940&      RG& 16 41 49.66& 36 27 48.6\cr
  v41&   782&      RG& 16 41 45.67& 36 27 57.4\cr
  v42&   289&      RG& 16 41 35.49& 36 27 27.2\cr
  v43&    96&      RG& 16 41 27.08& 36 28 00.2\cr
  v44&   445&      RG& 16 41 39.15& 36 27 21.9\cr
  v45&   554&      RG& 16 41 41.09& 36 27 22.7\cr
   \noalign{\vskip3pt\hrule\vskip3pt}
  v3&   135&   const& 16 41 29.75& 36 28 07.2\cr
  v4&   322&   const& 16 41 36.17& 36 28 51.5\cr
  v10&   487&   const& 16 41 39.96& 36 26 41.3\cr
  v12&   187&   const& 16 41 31.54& 36 28 44.1\cr
  v13&   327&   const& 16 41 36.65& 36 27 20.8\cr
  v21&   216&   const& 16 41 33.43& 36 27 10.7\cr
  v22&   568&   const& 16 41 41.27& 36 28 13.0\cr
  v23&   575&   const& 16 41 41.60& 36 27 38.4\cr
  v26&   748&   const& 16 41 45.08& 36 27 37.2\cr
  v27&   254&   const& 16 41 34.50& 36 27 52.3\cr
  v28&   519&   const& 16 41 40.44& 36 27 29.9\cr
  v29&   717&   const& 16 41 44.27& 36 28 31.7\cr
  v30&   743&   const& 16 41 44.97& 36 27 04.0\cr
   --&   222&   const& 16 41 33.64& 36 26 07.6\cr
   --&   261&   const& 16 41 34.74& 36 27 59.3\cr
   --&   334&   const& 16 41 36.77& 36 26 30.1\cr
    b&    --&   const& 16 41 49.79& 36 28 20.9\cr
    c&   993&   const& 16 41 52.05& 36 26 28.8\cr
  \noalign{\vskip3pt\hrule}}
  }\hfill}
 \end{table}

 \subsection{The case of v36}

 RRc variable v36 exhibits much larger
 scatter in the light curve
 than other variables of this type 
 (see Fig.~\ref{FigLCKopRRAll}).
 A closer look at the individual nights
 indicates modulation of the amplitude on time scale
 of several days. 

 \begin{figure}
  \hbox to\hsize{\hss\includegraphics{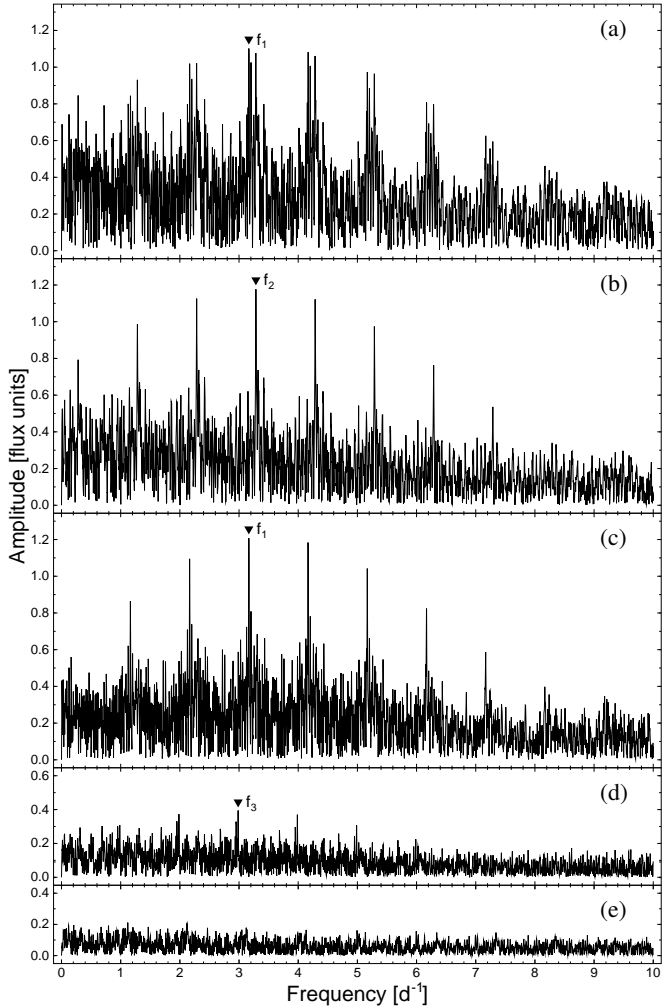}\hss}
  \caption{Fourier periodograms of the RRc variable v36:
   {\bf a}) for original $V$-filter data,
   {\bf b}) after prewhitening with frequency $f_1={}$3.166 d$^{-1}$,
   {\bf c}) after removing frequency $f_2={}$3.285 d$^{-1}$ only, 
   {\bf d}) after prewhitening with both $f_1$ and $f_2$ frequencies, and
   {\bf e}) after prewhitening with $f_1$, $f_2$, and $f_3={}$2.985 d$^{-1}$.
   Ordinates are expressed in the same flux units as in
   Fig.\ \ref{FigLCKopRRAll}.}
  \label{FigBPfourier}%
 \end{figure}

\begin{figure}
  \hbox to\hsize{\hss\includegraphics{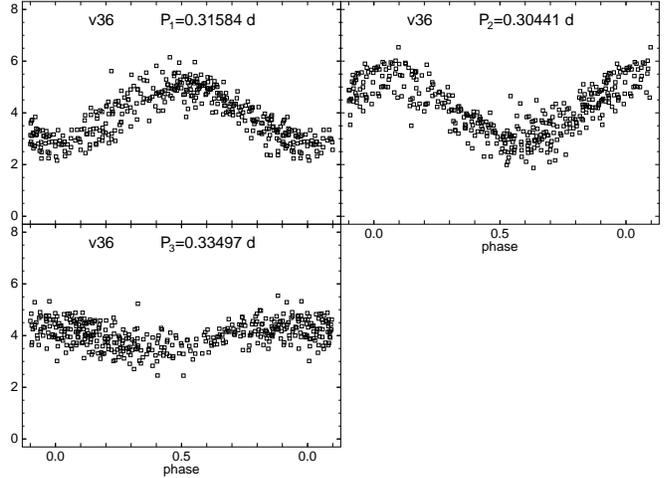}\hss}
  \caption{$V$-filter light-curves of the RRc variable v36.
   The upper left panel shows the data prewhitened with 
   $f_2$ and $f_3$ and phased with the period $P_1$.
   The upper right panel shows the residuals obtained after removing the
   $f_1$ and $f_3$ components, and phased with the period $P_2$.
   The lower panel shows the data prewhitened with 
   $f_1$ and $f_2$ and phased with the period $P_3$.
   Initial epoch was chosen arbitrarily, but it
   is the same for all light curves.
   Ordinate is expressed in the same flux units as in
   Fig.\ \ref{FigLCKopRRAll}.}
  \label{FigBPlc}%
 \end{figure}

 The Fourier spectrum of the original $V$-filter data 
 of v36 is presented in Fig.~\ref{FigBPfourier}a. 
 Two close peaks with almost the same height are seen in 
 this figure. The highest peak occurs at
 frequency $f_1={}$3.1661 d$^{-1}$
 ($P_1={}$0.31584 d). The periodogram of the 
 $I_{\rm C}$-filter observations (not given here)
 shows the same pattern of highest peaks.
 Periodogram of the residuals obtained
 after prewhitening the $V$-filter observations
 with the main frequency $f_1$ is shown in
 Fig.~\ref{FigBPfourier}b.
 The highest peak is seen at
 frequency $f_2={}$3.2850 d$^{-1}$ ($P_2={}$0.30441 d).
 In Fig.~\ref{FigBPfourier}d we show
 the periodogram of the $V$-filter residuals
 obtained after removing the $f_1$ and $f_2$
 frequencies.
 As can be seen, another significant
 peak appears at frequency
 $f_3={}$2.9853 d$^{-1}$ ($P_3={}$0.33497 d).
 It should be noted that this frequency is almost 
 equal to 3 d$^{-1}$ and at present
 we cannot be sure that it is real.
 After prewhitening the data with the
 $f_1$, $f_2$, and $f_3$, we obtain
 residuals for which a power spectrum is practically flat
 (Fig.~\ref{FigBPfourier}e).
 Prewhitened light-curves phased with periods $P_1$,
 $P_2$, and $P_3$, respectively, are shown in 
 Fig.~\ref{FigBPlc}.

 Thus, for v36 we get three close frequencies with a small
 separations, 
 $f_2-f_1 ={}$0.12 d$^{-1}$ and
 $f_1-f_3 ={}$0.18 d$^{-1}$.
 Our conclusion is that v36 belongs to the new group of
 the multi-mode RRc stars 
 discovered by Olech et al.\ (\cite{olechal99a}, \cite{olechal99b}).
 The frequencies of these variables
 are so closely spaced that the ratios of periods 
 are higher than 0.95 and could be as large as 0.999
 (Alcock et al.\ \cite{alcockal00}).
 For ratios of this order it is clear that
 if the main period is the first overtone radial
 mode then the other ones are non-radial modes. 
 However, in the case of v36
 it is not possible to unambiguously indicate
 which frequency corresponds to the radial
 mode, because $f_1$ and $f_2$ have almost the same
 amplitude. 
 Among the RRc stars in M\,13, the periods of v36 are
 the shortest ones. 

 At present, nine RR Lyrae stars 
 for which non-radial mode(s) have been detected
 are known in globular clusters;
 they are listed in Table~\ref{TabRRnonrad}.

 \begin{table}
  \caption[]{RR Lyrae stars in globular clusters
   for which non-radial modes have been found.}
  \label{TabRRnonrad}
  \hbox to\hsize{\hfill\vbox{\tabskip=4pt 
  \halign{#\hfil\tabskip=11pt&\hfil#&#\hfil&#\hfil\tabskip=3pt\cr
   \noalign{\hrule\vskip1.7pt}
   \noalign{\hrule\vskip3pt}
    Cluster& [Fe/H]& Variables& Source\cr
   \noalign{\vskip3pt\hrule\vskip3pt}
    M\,5&       $-$1.25& v104&         Olech et al.\ (\cite{olechal99b})\cr
    M\,13&      $-$1.65& v36&          this paper\cr
    M\,55&      $-$1.90& v9, v10, v12& Olech et al.\ (\cite{olechal99a})\cr
    M\,92&      $-$2.24& v11&          Kopacki (\cite{kopacki01})\cr
    NGC\, 6362& $-$1.08& v6, v10, v37& Olech et al.\ (\cite{olechal01})\cr
   \noalign{\vskip3pt\hrule}}
   }\hfill}
 \end{table}

 \section{Light-curve parameters}

 Mean brightnesses of variable stars were derived both
 as the intensity- and magnitude-weighted averages
 in the same way as in Kopacki (\cite{kopacki01}).
 However, instead of using DAOPHOT light-curves
 we used ISM light-curves transformed to the
 magnitude scale. The conversion of the ISM
 light-curves was accomplished by using all
 available DAOPHOT magnitudes. For each
 variable star the following equation
 $$m_i-m_0=-2.5\log{}(1+\Delta f_i/f_0) $$
 was solved by the method of the least-squares
 to find two parameters, reference point
 in magnitude, $m_0$, and reference point in the flux, $f_0$.
 Above, $m_i$ denote DAOPHOT magnitudes and
 $\Delta f_i$ ISM relative fluxes.
 Knowing $m_0$ and $f_0$ we transformed
 ISM light-curves into instrumental magnitudes
 and subsequently into the standard magnitudes.
 In this way, the quality of the derived light-curves
 (measured with the standard deviation of the
 magnitudes from the best finite Fourier series fit)
 was on average improved by a factor of 1.6.
 It should be mentioned that most authors
 derive reference points only from one frame, usually
 ISM reference image.
 However, it was already pointed out by Benk\H{o} (\cite{benko01}) that
 in order to obtain reliable transformations of the ISM
 fluxes into magnitudes, the reference magnitudes
 have to be known with a high accuracy.

 \begin{table}
  \caption[]{Average intensity-weighted,
   $\langle{V}\rangle_{\rm i}$,
   and magnitude-weighted, $\langle{V}\rangle_{\rm m}$, $V$ brightnesses,
   and the ranges of variability, $\Delta{V}$,
   for seven RR Lyrae stars and three BL Herculis
   variables, all located well outside
   the cluster core.
   Additionally, $\Delta{V}$ range, derived from the ISM
   photometry is shown for two other RR Lyrae stars,
   v25 and v35, situated in the central area
   of the cluster.
   $\sigma_V$ denotes the r.m.s.\ error of the average
   brightness derived from the best fit
   Fourier decomposition of the light curve.}
  \label{TabVarPhot}
  \hbox to\hsize{\hfill\vbox{\tabskip=5pt 
  \halign{#\hfil\tabskip=18pt%
    &\hfil#\hfil&\hfil#\hfil%
    &\hfil#\hfil&\hfil#\hfil\tabskip=3pt&\hfil#\hfil\cr
   \noalign{\hrule\vskip1.7pt}
   \noalign{\hrule\vskip3pt}
   Var& 
   $\langle{V}\rangle_{\rm i}$&
   $\langle{V}\rangle_{\rm m}$&
   $\Delta{V}$&
   $\sigma_V$&\cr
   \noalign{\vskip1pt}
      & [mag]& [mag]& [mag]& [mag]\cr
   \noalign{\vskip3pt\hrule\vskip3pt}
  v1&    14.086&   14.138&    1.041&    0.001\cr
  v2&    13.012&   13.054&    0.870&    0.001\cr
  v5&    14.767&   14.777&    0.425&    0.002\cr
  v6&    14.078&   14.095&    0.600&    0.001\cr
  v7&    14.925&   14.930&    0.305&    0.001\cr
  v8&    14.850&   14.880&    0.836&    0.001\cr
  v9&    14.819&   14.829&    0.434&    0.002\cr
  v31&   14.432&   14.433&    0.113&    0.001\cr
  v34&   14.828&   14.834&    0.320&    0.001\cr
  v36&   14.809&   14.810&    0.056\hbox to 0pt{ $f_1$\hss}&    0.001\cr
     &         &         &    0.064\hbox to 0pt{ $f_2$\hss}&    0.001\cr
     &         &         &    0.024\hbox to 0pt{ $f_3$\hss}&    0.001\cr
  \noalign{\vskip3pt\hrule\vskip3pt}
  v25&      --&       --&     0.44\blank&  0.01\blank\cr 
  v35&      --&       --&     0.25\blank&  0.01\blank\cr 
  \noalign{\vskip3pt\hrule}}
  }\hfill}
 \end{table}

 The light-curve parameters
 are given in Table~\ref{TabVarPhot}.
 Only variable stars for which a reliable DAOPHOT 
 profile photometry could be obtained are included
 in this table. For v36 we give full-amplitudes
 of all three detected modes.
 As can be seen in Table~\ref{TabVarPhot},
 the RR Lyrae star v31
 is about 0.4 mag brighter than the 
 other RR Lyrae variables. 
 We supposed that it is most probably caused by the presence 
 of a close companion unresolved in our CCD frames.
 Our suspicion has been confirmed by investigation of
 the recent HST pictures of M\,13, where v31
 is clearly seen as a close pair of stars.

 The mean $V$ magnitude of the horizontal branch,
 $\langle{V_{\rm HB}}\rangle_{\rm i}$,
 estimated using intensity-weighted
 mean magnitudes of six RR Lyrae
 stars listed in Table~\ref{TabVarPhot}
 (we excluded v31 for the reason given above),
 is equal to $14.83\pm0.02$ mag.

 As discussed
 in detail by Kopacki (\cite{kopacki01}) and
 Benk\H{o} (\cite{benko01}),
 the ISM of Alard \&\ Lupton (\cite{alardlupton98})
 does not allow one to derive light curves
 expressed in magnitudes without knowing
 at least one reference magnitude corresponding
 to some ISM relative flux. This additional
 parameter is usually derived from the DAOPHOT
 photometry and is lacking for stars located
 in the crowded core of the globular cluster.
 However, we expect all RR Lyrae stars in the
 cluster to have very similar average brightness,
 and this average brightness may be used
 as a reference magnitude, although exclusively
 for RR Lyrae variables.

 Using several RR Lyrae stars for which
 both reliable DAOPHOT and ISM light-curves
 were obtained we determined
 linear transformation
 between the flux measured by DAOPHOT
 and differential flux of the ISM.
 Assuming that RR Lyrae stars lying
 in the cluster central region have
 the same intensity-weighted average brightness,
 equal to the average magnitude of the
 horizontal branch, $\langle{V_{\rm HB}}\rangle_{\rm i}$,
 we were able to compute $\Delta{V}$ ranges 
 of these stars.
 $\Delta{V}$ ranges of
 the two RR Lyrae variables, v25 and v35,
 derived in this way, 
 are listed in Table~\ref{TabVarPhot}.

\begin{figure}
  \hbox to\hsize{\hss\includegraphics{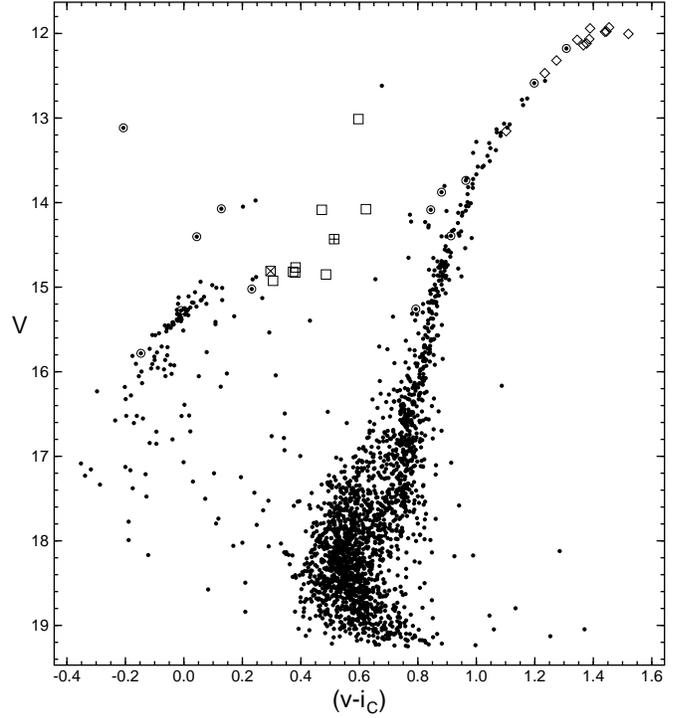}\hss}
  \caption{The $V$ vs.\ $(v-i_{\rm C})$ colour-magnitude
  diagram for M\,13. Only stars with the
  distance from cluster centre greater than 1.8 arcmin
  and the error in $(v-i_{\rm C})$ smaller than 0.1 mag are included.
  RR Lyrae and BL Herculis stars are shown with open squares
  and are represented by their intensity-weighted
  mean $V$ brightness and magnitude-weighted mean
  $(v-i_{\rm C})$ colour index.
  Multi-periodic RRc star v36 is additionally marked with the `x' sign.
  Semiregular variable red giants are plotted
  as open diamonds.
  Some suspected variable stars listed
  in the CVSGC, which turned out to be constant, are
  also shown with symbols enclosed within circles.
  Note the unusual position of an RRc variable v31
  (indicated with the plus sign), 
  caused by the presence of a close companion unresolved
  in our CCD frames.}
  \label{FigCMD}%
 \end{figure}

\section{The colour-magnitude diagram}

 The $V$ vs.\ $(v-i_{\rm C})$ colour-magnitude 
 diagram for M\,13 is shown in
 Fig.~\ref{FigCMD}. Only stars with
 distance from the cluster center greater than 1.8 arcmin
 and having an error in the colour-index smaller than 0.1 mag are plotted.
 Variable stars are represented by their intensity-weighted
 mean $\langle{V}\rangle_{\rm i}$ brightness and 
 instrumental magnitude-weighted mean 
 $\langle{v-i_{\rm C}}\rangle_{\rm m}$ colour index.

 The colour-magnitude diagram shows a well-defined
 red giant branch, 
 asymptotic giant branch, and
 the predominantly blue horizontal branch. 
 There is a clear separation between the variable and 
 constant stars and also between one RRab star and RRc variables.
 One should note an unusual position of the
 RRc variable v31, which is brighter and redder than
 all other RR Lyrae stars in the cluster. As we already pointed out,
 it is caused by blending with another
 star.

 In the colour-magnitude diagram the multi-periodic non-radial
 pulsating RRc variable v36 is situated at the blue
 edge of the instability strip. The same result was
 obtained by
 Olech et al.\ (\cite{olechal99a}) for three
 non-radial pulsating RRc stars in M\,55,
 by Olech et al.\ (\cite{olechal01}) 
 in the case of also three RRc variables 
 in the cluster NGC\,6362, and
 by Kopacki (\cite{kopacki01}) for one such
 variable in M\,92.

 Thirteen suspected variable stars for which 
 reliable photometry was obtained but
 did not show evidence of variations
 are also shown in Fig.~\ref{FigCMD}.
 Three of them, v4, v12, and v29, occupy the blue part of the
 horizontal branch and lie outside the
 RR Lyrae instability strip. 
 The bright blue star is L222, 
 the famous blue straggler also known as Barnard 29
 (Barnard \cite{barnard31}).
 v27 and L993 seem to be field stars,
 although L993 is a known UV-bright object
 classified as a cluster member on the basis of 
 radial velocity and proper motion measurements. 
 The remaining seven, v3, v10, v13, v21, v30, 
 L261, and L334,
 are red giants, with v13 and v21 being 
 most probably asymptotic branch objects.

\section{Summary}

We presented two-colour photometric study of 
variable stars in globular cluster M\,13.
The search for new variable stars resulted in
the discovery of three RR Lyrae stars and one 
SX Phoenicis variable. Since the
image subtraction method was used, our 
search is complete as far as RR Lyrae stars
are concerned. It appears that M\,13 is
relatively poor in RR Lyrae stars.
Apparently, it contains nine variables
of this type with only one being
an RRab star. Among RRc stars
we found one variable (v36) which pulsates
in at least two modes with closely-spaced frequencies.
It is therefore an RR Lyrae star with nonradial mode(s) excited.
Moreover, we found that all cluster red giants 
brighter than about 12.5 mag in $V$ are variable.
Only one variable red giant exhibits
periodic variations and all remaining ones
are semi- or irregular.

 \begin{acknowledgements}
  We want to express here our gratitude to 
  Prof.\ M.\ Jerzykiewicz for
  critical reading of the manuscript.
  We are also grateful to the referee
  for valuable comments.
  This work was supported by the KBN grant No.\ 2 P03D$\,$006$\,$19.
 \end{acknowledgements}

\end{document}